\documentstyle[11pt]{article}
\def\fnote#1#2{\begingroup\def\thefootnote{#1}\footnote{#2}\endgroup}
\def\BM#1{\mbox{\boldmath{$#1$}}}
\begin{document}

\hfill{UTTG-03-01}

\vspace{36pt}

\begin{center}
{\large{\bf { Fluctuations in the Cosmic Microwave Background I: Form Factors 
and their Calculation in Synchronous Gauge }}}

\vspace{36pt}
Steven Weinberg\fnote{*}{Electronic address:
weinberg@physics.utexas.edu}\\
{\em Theory Group, Department of Physics, University of
Texas\\
Austin, TX, 78712}
\end{center}

\vspace{30pt}

\begin{center}
{\bf Abstract}
\end{center}

\noindent
It is shown that the fluctuation in the temperature of the cosmic microwave 
background in any direction may be evaluated as an integral involving scalar and 
dipole form factors, which incorporate all relevant information about acoustic 
oscillations before the time of last scattering. A companion paper gives 
asymptotic expressions for the multipole 
coefficient $C_\ell$ in terms of these form factors.
Explicit expressions are given here
for the form factors in a simplified hydrodynamic model for the evolution of  
perturbations.  
\vfill

\pagebreak
\setcounter{page}{1}

\begin{center}
{\bf I. INTRODUCTION}
\end{center}

The purpose of this paper is, first,  to exhibit a general formalism, expressing 
the observed fluctuations in the cosmic microwave background temperature in 
terms of a pair of form factors, and then to carry out an illustrative 
approximate analytic 
calculation of these form factors.  A companion paper[1]  gives general 
asymptotic formulas for the coefficient $C_\ell$ of the $\ell$th multipole term 
in the temperature correlation function for arbitrary form factors, and also 
uses these formulas to calculate $C_\ell$ for the form factors found in the 
present paper.

In Section II we show that under very general assumptions the fractional 
variation from the mean of the cosmic microwave background temperature observed 
in a direction $\hat{n}$ takes the form
\begin{equation}
\frac{\Delta T(\hat{n})}{T} 
=\int d^3k\,\epsilon_{\bf k}\,e^{id_A\hat{n}\cdot{\bf 
k}}\,\left[F(k)+i\,\hat{n}\cdot\hat{k}\,G(k)\right]\;.
\end{equation}
Here $d_A$ is the angular diameter distance of the surface of 
last scattering,\fnote{**}{ Note that in speaking of a surface of last 
scattering, we are not necessarily assuming that the transition from opacity to 
transparency takes place instantaneously.   The physical wave number vector 
${\bf k}$ varies with time as $1/a(t)$ (where $a(t)$ is the 
Robertson--Walker scale factor), while for large redshifts $d_A$ varies 
as $a(t)$, so the product $d_A{\bf k}$ is nearly independent  of what we choose 
as a nominal time of last scattering.} and ${\bf k}^2\epsilon_{\bf k}$ is 
proportional (with a ${\bf k}$-independent proportionality coefficient) to the 
Fourier transform of the 
fractional perturbation in the total energy density at early times.   (There are 
additional terms in $\Delta T/T$ that arise from times near the present, and 
chiefly effect the multipole coefficients $C_\ell$ for small $\ell$, especially 
$\ell=0$ and $\ell=1$.  
These will be discussed in Section IV and in the Appendix.  Effects from a 
changing gravitational field soon after the time of last scattering are included 
in Eq.~(1).)

One advantage of this formalism is that it  provides a nice separation between 
the three different kinds of 
effect that influence the observed temperature fluctuation, that arise in three 
different eras: (1) at very early times, (2) during the era of acoustic 
oscillations, and (3) from the time of last scattering to the present:

\begin{enumerate}
\item The  ${\bf k}$-dependence of the unprocessed fluctuation amplitude 
$\epsilon_{\bf k}$ 
reflects the space-dependence of  fluctuations in the energy density at very 
early times.  The average of the product of two $\epsilon$s is assumed to 
satisfy the conditions of statistical homogeneity and isotropy:
\begin{equation} 
\left\langle \epsilon_{\bf k}\,\epsilon_{\bf k'}\right\rangle
=\delta^3({\bf k}+{\bf k'})\,{\cal P}(k)
\end{equation} 
with $k\equiv |{\bf k}|$. Since the reality of the fluctuations in the energy 
density requires that $\epsilon_{\bf k}^*=\epsilon_{-{\bf k}}$, the 
power spectral function ${\cal P}(k)$ is real and positive.  It is common to 
assume a ``straight'' spectrum
\begin{equation}\
{\cal P}(k)\propto k^{n-4}
\end{equation} 
For instance, the `scale-invariant' $n=1$ form[2] suggested by theories of new 
inflation[3] is: 
\begin{equation}
{\cal P}(k)=B\,k^{-3}\;,
\end{equation} 
with $B$ a constant that must be taken from observations of the cosmic microwave 
background or condensed object mass distributions, or from detailed theories of 
inflation. 

\item
The form factors $F(k)$ and $G(k)$ characterize acoustic oscillations, with 
$F(k)$ arising from the Sachs--Wolfe effect and intrinsic temperature 
fluctuations, and $G(k)$ arising from the Doppler effect.

\item
Taking account of a constant vacuum energy density as well as cold matter in the 
time after last scattering, it is easy to calculate the  angular diameter 
distance of the surface of last scattering:
\begin{equation}
d_A=\frac{1}{\Omega_C^{1/2}H_0(1+z_L)}\sinh\left[\Omega_C^{1/2}\int_{\frac{1}{1+
z_L}}^1\frac{dx}{\sqrt{\Omega_\Lambda x^4+\Omega_Cx^2+\Omega_Mx}}\right]\;,
\end{equation} 
where $z_L\simeq 1100$ is the redshift of last scattering, $\Omega_C\equiv1-
\Omega_\Lambda-\Omega_M$, and $\Omega_\Lambda$ and $\Omega_M$ are as usual the 
present 
ratios of the energy densities of the vacuum and matter to the critical density 
$3H_0^2/8\pi G$.
\end{enumerate}
We note in particular that $F(k)$ and $G(k)$ depend on $\Omega_M h^2$ and on the 
baryon density parameter $\Omega_B h^2$ (where $h$ is the Hubble constant in 
units of 100 km/sec/Mpc), but since the curvature and vacuum 
energy were negligible at and before last scattering, $F(k)$ and $G(k)$ are 
essentially independent of the present curvature and of $\Omega_\Lambda$.  The 
exponent $n$ in ${\cal P}(k)$ is expected to be independent of all these 
parameters.  On the other hand, $d_A$ is affected by whatever governed the paths 
of light rays since the time of last scattering,
so it depends on $\Omega_M$, $\Omega_\Lambda$, and the curvature parameter 
$\Omega_C$, but it is essentially independent of quantities like the baryon 
density parameter $\Omega_B$ that effect acoustic oscillations before the time 
of last scattering.  In quintessence theories $d_A$ would be given by a 
different formula, but ${\cal P}(k)$ and the form factors would be essentially 
unchanged as long as the quintessence energy density is a small part of the 
total energy density at and before the time of last scattering.

Another advantage of this formalism is that, although $C_\ell$ must be 
calculated by a numerical integration, it is possible to give approximate 
analytic expressions for the form factors in terms of elementary functions.  The 
detailed confrontation of observation and theory must necessarily be done 
using computer codes that take into account all relevant astrophysical and 
observational effects.[4]  Nevertheless, there is some value in also having an 
analytic treatment that, though not as accurate as possible, is as simple as 
possible while still capturing the main features of what is going on.  The 
point is not to compete with the computer codes, but rather to gain some feeling 
for what is going on, in order to help us judge how predictions for the cosmic 
microwave background fluctuations may change with alterations in the underlying 
assumptions. 

Analytic treatments of fluctuations in the cosmic microwave 
background already exist in the literature[5].  Our main purpose in going over 
the same ground here is not to give a more accurate or comprehensive treatment 
of acoustic oscillations, but to obtain simple expressions for the form factors 
as examples to which to apply the asymptotic formulas for $C_\ell$ derived in 
reference [1].  To derive analytic expressions for the temperature fluctuation 
it is necessary to neglect the contribution of radiation and neutrinos to the 
gravitational field, which should be a fair approximation near the first Doppler 
peak but not much beyond that.  We employ a purely hydrodynamic treatment, 
relying on the Boltzmann equation only implicitly in the values used for the 
shear viscosity and heat conduction coefficients; the effects of viscosity and 
heat conduction are included from the beginning, not just by inserting damping 
factors; and `Landau' damping due to the finite duration of the era of last 
scattering is included along with `Silk' damping due to shear viscosity and heat 
conduction.  As far as I know, this is the first work to obtain {\em explicit} 
analytic expressions for the temperature fluctuations that are correct within 
these approximations.  

Section III presents an analytic calculation of the evolution of 
perturbations in synchronous gauge up to the time of last scattering, which is 
then used in Section IV to calculate the form factors. For very small wave 
numbers the form factors are found to be
\begin{eqnarray} 
F(k)&\rightarrow & 1-3k^2t_L^2/2-3[-\xi^{-
1}+\xi^{-2}\ln(1+\xi)]k^4t_L^4/4+\dots\;,\\
G(k)&\rightarrow & 3kt_L-3k^3t_L^3/2(1+\xi)+\dots\;,
\end{eqnarray} 
while for wave numbers large enough to allow the use of the WKB approximation, 
i. e., $kt_L\geq \xi$, 
the form factors are[6]
\begin{equation}
F(k)= (1+2\xi/k^2t_L^2)^{-1}\left [-3\xi+2\xi/k^2t_L^2+(1+\xi)^{-1/4}e^{-
k^2d_D^2}\cos(kd_H)\right]\;,
\end{equation} 
 and 
\begin{equation}
G(k)=\sqrt{3}\,(1+\xi)^{-3/4}(1+2\xi/k^2t_L^2)^{-1} e^{-
k^2d_D^2} \sin(kd_H)\;.
\end{equation}
Here $t_L$ is the time of last scattering; $\xi$ is $3/4$ the ratio of the 
baryon to photon energy densities at this time:
\begin{equation}
\xi=\left(\frac{3\rho_B}{4\rho_\gamma}\right)_{t=t_L}\simeq 27\,\Omega_Bh^2\;;
\end{equation} 
$d_H$ is the acoustic horizon size at this time, given by Eq.~(75), and 
$d_D $ is a damping length, given by Eq.~(89).

\begin{center}
{\bf II. FORM FACTORS}
\end{center}

We first justify Eq.~(1) under very general assumptions, not limited to those of 
Section III.  At and before the time of last scattering the spatial 
curvature was negligible, 
so small perturbations in the cosmic metric and in all particle distributions at 
these times may conveniently be expressed as Fourier transforms of functions of 
a 
co-moving wave number vector ${\bf q}$ and the time $t$.  Effects like pressure 
forces that involve spatial gradients are important for a given ${\bf q}$ only 
when the physical wave number $q/a(t)$ is at least as large as the cosmic 
expansion rate, which is of order $1/t$.  Since $a(t)$ vanishes for 
$t\rightarrow 0$ no more rapidly than $\sqrt{t}$, the ratio $qt/a(t)$ vanishes
as $t\rightarrow 0$, so whatever the value of $q$, there will always be some 
time early enough so that pressure forces and other effects of spatial gradients 
are negligble.  At such early times, perturbations grow or decay with powers of 
time.  Generically  there is one most 
rapidly growing mode, and this  is the one that eventually grows into the 
perturbations seen at the time of last scattering.  
Since the equations for the time dependence of the perturbations are linear, the 
Fourier transforms of all perturbations to the metric and particle distributions 
during the era of last scattering will then be proportional to the Fourier 
transform $e_{\bf q}$ of any one of these perturbations at any sufficiently 
early time.  For definiteness, we can take $e_{\bf q}$ to be $q^{-2} $ times 
the Fourier transform of the fractional perturbation to the total energy density 
at some very early time, a choice that will prove` to be convenient in Section 
IV.  

Since the fractional change in the observed microwave background temperature 
seen in a direction $\hat{n}$ is linear in the perturbations to the metric and 
photon and matter distributions at various times during the era of last 
scattering, it can be written as
\begin{equation}
\frac{\Delta T(\hat{n})}{T}=\int dt\int d^3q\;e^{i{\bf q}\cdot \hat{n}r(t)}
\,e_{\bf q}\,J(q, \hat{q}\cdot\hat{n}, t) \;,
\end{equation} 
where $r(t)$ is the co-moving radial coordinate of a source scattering light at 
time $t$ that would be received at the present.
Note that the quantity $J$ can depend on ${\bf q}$ only through the scalars $ q$ 
and $\hat{q}\cdot\hat{n}$, because the differential equations governing the 
growth of perturbations are rotationally invariant, even though the initial 
fluctuation amplitude $e_{\bf q}$ is not.   

We can make a great simplification in Eq.~(11) by taking advantage of the fact 
that the Robertson--Walker radial coordinate $r(t)$ is nearly constant during 
the era of last scattering.  Using equilibrium statistical mechanics to 
calculate the hydrogen ionization, and simple Thomson scattering to calculate 
scattering probabilities, one finds that for $\Omega_B/\Omega_M=0.2$, the 
probability that a photon will 
never again be scattered rises from 2\% at 3360$^\circ$K to 98\% at 
2780$^\circ$K.  (This   depends on the assumed value of $\Omega_B/\Omega_M$, but 
very weakly; for instance, for $\Omega_B/\Omega_M=0.12$, the probability of no 
future scattering rises from 2\% to 98\% as the temperature drops from 
3400$^\circ$K to 2810$^\circ$K.)   For definiteness, we will round off these 
temperatures, taking the era of last scattering to extend from a temperature of 
3400$^\circ$K  down to 2800$^\circ$K, corresponding to a drop in redshift $z$ 
from 1220 to 1010.  The 
radial coordinate can be expressed in terms of $z$ by the well-known formula
\begin{equation}
r(t_z)=\frac{1 }{\Omega_C^{1/2}H_0 
a(t_0)}\sinh\left[\Omega_C^{1/2}\int_{\frac{1}{1+z}}^1\frac{dx}{\sqrt{
\Omega_\Lambda x^4+\Omega_Cx^2+\Omega_Mx}}\right]\;,
\end{equation} 
where $\Omega_C\equiv 1-\Omega_\Lambda-\Omega_M$, $a(t)$ is  the 
Robertson--Walker scale factor, $t_z$ is the time corresponding to redshift $z$, 
and $t_0$ is the present.
This approaches a constant limit for $z\rightarrow \infty$, and therefore varies 
very little in the range from $z=1010$ to $z=1220$ (or, for that matter, even in 
the range from $z=1010$ to $z\rightarrow \infty$.)  For instance, if we take the 
popular values $\Omega_M=0.3$ and $\Omega_V=0.7$, then the fractional change in 
$r(t_z)$ as $z$ drops from 1220 to 1010 is 0.0034.  We can therefore take the 
exponential in Eq.~(11) outside the time integral, so that 
\begin{equation}
\frac{\Delta T(\hat{n})}{T}= \int d^3q\;e^{i{\bf q}\cdot \hat{n}\,r(t_L)}\, 
e_{\bf q}\int 
dt 
\,J(q, \hat{q}\cdot\hat{n}, \,t) \;,
\end{equation} 
where $t_L$ is any conveniently chosen time during the era of last scattering, 
say at a redshift $z_L=1100$.  

It is convenient to replace the co-moving wave number vector ${\bf q}$ with the 
physical wave number at last scattering:
\begin{equation}
{\bf k}\equiv {\bf q}/a(t_L)\;.
\end{equation}  Eq.~(13) may then be written
\begin{equation}
\frac{\Delta T(\hat{n})}{T}= \int d^3k\;e^{i{\bf k}\cdot 
\hat{n}d_A}\,\epsilon_{\bf k} \int dt 
\,{\cal J}(k, \hat{k}\cdot\hat{n}, t) \;,
\end{equation} 
where ${\cal J}(k, \hat{k}\cdot\hat{n}, t)\equiv a(t_L)^3J(q, 
\hat{q}\cdot\hat{n}, t)$, $\epsilon_{\bf k}\equiv e_{\bf q}$, and $d_A\equiv 
r(t_L)a(t_L)$ is the angular diameter distance of the surface of last 
scattering.

Part of the observed temperature fluctuations arise from perturbations in scalar 
quantities, like the gravitational potential and the intrinsic temperature,
and therefore makes a contribution to ${\cal J}$ that is independent of 
$\hat{n}$.
Another part arises from fluctuations in a vector, the velocity of the 
baryon--electron plasma, and  therefore makes a contribution that is linear in 
$\hat{n}$.  Leaving aside other effects like gravitational radiation, the 
function ${\cal J}$ therefore takes the form
\begin{equation}
{\cal J}(k, \hat{k}\cdot\hat{n}, \,t)= {\cal F}(k,t) 
+i\hat{k}\cdot\hat{n}\,{\cal G}(k,t)\;,
\end{equation} 
with $ {\cal F}(k,t)$ arising from the Sachs--Wolfe effect and intrinsic 
temperature fluctuations, and $ {\cal G}(k,t)$ arising from the Doppler effect.
Using this in Eq.~(15) then gives
\begin{equation}
\frac{\Delta T(\hat{n})}{T}= \int d^3k\;e^{i{\bf k}\cdot \hat{n}d_A}  
\,\left[F(k)+i\hat{k}\cdot\hat{n}\,G(k)\right]\,\epsilon_{\bf k}\;,
\end{equation}
which is the same as Eq.~(1), with the form factors identified as time integrals
\begin{equation}
F(k)\equiv \int dt\;{\cal F}(k,t)\:,~~~~~~~~~~~~ G(k)\equiv \int dt\;{\cal 
G}(k,t)\;.
\end{equation} 
This time integration introduces a damping of oscillatory part of the form 
factors, but this will be less important than the effects of heat conduction and 
viscosity in the time interval between recombination and last scattering.

\begin{center}
{\bf III. EVOLUTION OF PERTURBATIONS IN SYNCHRONOUS GAUGE }
\end{center}

We now turn to the approximate analytic calculation of the form factors.

\begin{center}
{\bf A. General Approximations}
\end{center}

We 
make two assumptions that will allow great simplifications in this calculation:
\begin{enumerate}
\item The contents of the universe up to the time of last scattering are taken 
to consist of collisionless cold dark matter, collisionless neutrinos, a 
baryon--electron plasma treated as a perfect fluid, and a photon gas coupled to 
the plasma by Thomson scattering, with a short but non-negligible photon mean 
free time.  The finite duration of the era of last scattering, when the mean 
free time becomes too large to allow a hydrodynamic treatment,  will be taken 
into account by the time integrals in Eq.~(18).
\item It is assumed that only the cold dark matter contributes to the expansion 
rate of the universe before the time of last scattering and to perturbations in 
the metric.  This is not a very good approximation, but it is the price that has 
to be paid to get analytic expressions for the observed temperature fluctuation.  
To minimize errors introduced by the incorrect treatment of acoustic 
oscillations before the cross-over time $t_C$ 
when the photon energy density equaled  the dark matter energy density, it is 
necessary to restrict the wave number to be less than an upper bound given in 
Section V.
\end{enumerate}

\begin{center}
{\bf B. Gravitational Field}
\end{center}

We begin by reminding the reader of the equations that govern perturbations in 
the metric and fluid properties before the time of last scattering.  The 
perturbed metric is taken as
\begin{equation}
{g}_{\mu\nu\;{\rm total}}({\bf x},t)= g_{\mu\nu}(t)+h_{\mu\nu}({\bf x},t)\;,
\end{equation} 
where $g_{\mu\nu}$ is the Robertson--Walker metric in co-moving coordinates 
${\bf x}$ with spatial curvature neglected:
\begin{equation}
g_{00}=-1\;,\qquad g_{0i}=0\;,\qquad g_{ij}=a^2(t)\delta_{ij}\;,
\end{equation}
and $h_{\mu\nu}({\bf x},t)$ 
is a small perturbation. We work in synchronous gauge, defined by the conditions
\begin{equation}
h_{0i}=h_{00}=0\;,
\end{equation} 
and by the requirement that the cold dark matter particles have time-independent 
spatial coordinates.
These conditions leave an unbroken residual gauge invariance, under the 
transformation
\begin{equation}
h_{ij}\rightarrow h_{ij}+a^2\left(\frac{\partial e_i}{\partial x_j}
+\frac{\partial e_j}{\partial x_i}\right)\;,
\end{equation} 
with $e_i$ an arbitrary  function of ${\bf x}$ but independent of $t$.  As we 
will see, in synchronous gauge the evolution of the compressional modes that 
concern 
us here depends on the gravitational field only through a quantity that is 
invariant under these these transformations
\begin{equation}
\psi\equiv \frac{\partial}{\partial t}\left(\frac{h_{kk}}{2a^2}\right)
\end{equation} 
The spatial curvature is negligible at and before the time of last scattering, 
so it will be convenient to express $\psi({\bf x}, t)$ as a Fourier transform:
\begin{equation} 
\psi({\bf x}, t)=\int d^3q\;e^{i{\bf q}\cdot{\bf x}}\,\psi_{\bf q}(t)\;.
\end{equation}
Likewise, the total proper energy density of each of the 
constituents of the universe (labelled $f=D,\,B,\,\gamma$ for dark matter, the 
baryon--electron plasma, and photons, respectively) is written
\begin{equation}  
\varrho_{f\;{\rm total} }({\bf x},t)=\varrho_{f}(t)+\delta\rho_f({\bf 
x},t)\;,~~~~~~~~~~~~ \delta\rho_f({\bf x},t)=\int d^3q\;\varrho_{f{\bf 
q}}(t)\;e^{i{\bf q}\cdot{\bf x}} \;,
\end{equation} 
with quantities carrying a subscript ${\bf q}$  denoting  small perturbations. 
Under our second assumption, the gravitational field equation in synchronous 
gauge reads[7]
\begin{equation}
\frac{d}{d t}\Big(a^2\psi_{\bf q}\Big)=-4\pi Ga^2\rho_{D{\bf q}}\;.
\end{equation}

\begin{center}
{\bf C. Dark Matter Perturbations}
\end{center}

The dark matter particles are assumed to ride on the expanding coordinate mesh, 
with negligble peculiar velocities.  (This is not affected by perturbations to 
the gravitational field, because in synchronous gauge these perturbations leave 
$\Gamma^i_{00}$ zero.)  Hence  their energy-momentum tensor has only a 
$00$-component, $T_D^{00}=\varrho_{D\;{\rm total}}$.  For the metric (19)--(21), 
the 
energy conservation equation ${T_D^{0\mu}}_{;\mu}=0$ then reads 
$$
\frac{d\rho_{D{\bf q}}}{dt}+\frac{3\dot{a}}{a}\rho_{D{\bf q}}+\psi_{\bf 
q}\rho_D=0
$$ 
or in other words
\begin{equation}
\frac{d \delta_{D{\bf q}}}{d t}=-\psi_{\bf q}\;,
\end{equation} 
where $\delta_{D{\bf q}}$ is the fractional dark matter density perturbation:
\begin{equation} 
\delta_{D{\bf q}}\equiv \rho_{D{\bf q}}/\rho_D\;.
\end{equation} 
Combining Eqs.~(26) and (27) gives
\begin{equation}
\frac{d}{d t}\left(a^2\frac{d \delta_{D{\bf q}}}{d t}\right)=4\pi 
Ga^2\rho_D\delta_{D{\bf q}}\;.
\end{equation} 
During the dark matter dominated era, $a\propto t^{2/3}$ and $4\pi G 
\rho_D=2/3t^2$,
so Eq.~(29) can be written
\begin{equation}
\frac{d}{d t}\left(t^{4/3} \frac{d \delta_{D{\bf q}}}{d 
t}\right)=\frac{2}{3}t^{-2/3} \delta_{D{\bf q}}\;.
\end{equation} 
As is well known, the two solutions go as $t^{-1}$ and $t^{2/3}$.  If these two 
modes have comparable strengths for very small $t$, then the relevant 
solution is the one that is most rapidly growing, which we shall write as
\begin{equation}
\delta_{D{\bf q}}=N_{\bf q}t^{2/3}\;,~~~~~~~~~~~\psi_{\bf q}= -\frac{2}{3}N_{\bf 
q}t^{-1/3}\;.
\end{equation} 
(The normalization constant $N_{\bf q}$ will play a role in this section similar 
to that of the constant $e_{\bf q}$ in Section II.)

\begin{center}
{\bf D. Plasma and Photon Perturbations}
\end{center}

Next, let us consider the imperfect fluid formed by the baryon--electron plasma 
and the photons.  It has a total velocity four-vector of the form 
\begin{equation}
U_{\rm total}^\mu({\bf x},t)=U^\mu+\int d^3q\;U_{\bf q}^\mu(t)\;e^{i{\bf 
q}\cdot{\bf x}}
\end{equation} 
where $U^\mu$ is the unperturbed velocity four-vector
\begin{equation}
U^0=1\;,~~~~~~~~U^i=0\;,
\end{equation}
and $U_{\bf q}^\mu(t)$ is a small perturbation. 
The normalization condition $g_{\mu\nu\;{\rm total}}U^\mu_{\rm total}
U^\nu_{\rm total}=-1$ tells us that the first-order perturbations are purely 
spatial
\begin{equation}
U_{\bf q}^0(t)=0\;.
\end{equation} 
We will be considering only compressional modes, so we will assume that the 
spatial part of $U^\mu_{\rm total}$ is the gradient of a velocity potential $u$:
\begin{equation}
U_{\bf q}^i(t)=i\,q^i u_{\bf q}(t)\;.
\end{equation} 
We will write the conservation laws for this fluid in terms of fractional 
perturbations to the baryon--electron plasma mass density and the photon energy 
density:
\begin{equation} 
\delta_{B{\bf q}}\equiv \rho_{B{\bf q}}/\rho_B\;,~~~~~~~~~~~~\delta_{\gamma{\bf 
q}}\equiv\rho_{\gamma{\bf q}}/\rho_\gamma\;.
\end{equation} 
The particle conservation equation[8] for the baryon--electron plasma  mass 
density is then 
\begin{equation}
\frac{d\delta_{B{\bf q}}}{d t}=q^2u_{\bf q}-\psi_{\bf q}\;.
\end{equation}
The energy conservation equation[9] for the baryon--electron--photon fluid is
\begin{eqnarray} 
&&\frac{d}{dt}\Big(\rho_B\delta_{B{\bf q}}+ \rho_\gamma\delta_{\gamma{\bf 
q}}\Big)+\frac{3\dot{a}}{a}\left(
\rho_{B}\delta_{B{\bf q}}+ \frac{4}{3}\rho_{\gamma}\delta_{\gamma{\bf 
q}}\right)= -\left(\rho_B+ \frac{4}{3}\rho_\gamma\right)\,\Big(\psi_{{\bf q}}-
q^2 u_{{\bf q}}\Big)
\nonumber\\&&~~~~~~~~~~~~~-\chi q^2\left(\dot{T}u_{{\bf q}}+\frac{T}{a^2}\frac{d 
(a^2u_{{\bf q}})}{d t}+\frac{T\delta_{\gamma{\bf q}}}{4a^2}\right)\;,
\end{eqnarray} 
where  $T$ is the unperturbed photon temperature and
$\chi$ is the coefficient of heat conduction caused by photon energy transport.
Finally, the momentum conservation equation[10] is 
\begin{eqnarray} 
&&\left[\frac{d}{d t}+16\pi G\eta\right]\left[-
a^5\left(\rho_B+\frac{4}{3}\rho_\gamma-\chi \dot{T}\right)\,u_{\bf q}+\chi T 
a^3\left(\frac{\delta_{\gamma{\bf q}}}{4}+\frac{d}{d t}\Big(a^2u_{{\bf 
q}}\Big)\right)\right]\nonumber\\
&&~~~~~~~~~~=\frac{1}{3}a^3 \rho_\gamma \delta_{\gamma{\bf q}}-\frac{4\eta 
a^3}{3}\left[-q^2u_{\bf q}+\psi_{{\bf q}}\right]\;,
\end{eqnarray}
where $\eta$ is the coefficient of viscosity due to photon momentum transport.
By using Eq.~(37) and recalling that $\rho_B\propto a^{-3}$, $\rho_\gamma\propto 
a^{-4}$, and 
$T\propto a^{-1}$, we can simplify Eq.~(38) to read
\begin{equation} 
 \frac{d}{dt}\left[\delta_{\gamma{\bf q}}-\frac{4}{3}\delta_{B{\bf q}}\right] 
=\frac{\chi T}{\rho_\gamma}\left[-\frac{1}{a}\frac{\partial}{\partial t} 
\Big(a(\dot{\delta}_{B{\bf q}}+\psi)\Big)-\frac{q^2\delta_{\gamma{\bf 
q}}}{4a^2}\right]\;,
\end{equation} 
Also, using Eq.~(37) lets us write Eq.~(39) as
\begin{eqnarray} 
&&\left[\frac{d}{d t}+16\pi 
G\eta\right]\left[a^5\left(\rho_B+\frac{4}{3}\rho_\gamma-\chi 
\dot{T}\right)\,\Big(\dot{\delta}_{B{\bf q}}+\psi_{{\bf q}}\Big)
\right. \nonumber\\
&&~~-\left.\chi T a^3\left(\frac{q^2\delta_{\gamma{\bf q}}}{4}+\frac{d}{d 
t}\Big(a^2(\dot{\delta}_{B{\bf q}}+\psi_{\bf q})\Big)\right)\right] =-
\frac{q^2a^3}{3}\left[\rho_\gamma \delta_{\gamma{\bf q}}+4\eta 
\dot{\delta}_{B{\bf q}}\right]\;.~~~~~~~~
\end{eqnarray}

Now, $\eta/\rho_\gamma$ and $\chi T/\rho_\gamma$ are of the order of the photon 
mean free time, which as long as hydrodynamics is applicable must be short 
compared with the cosmic age.  Therefore we can neglect $\eta$ and $\chi$ 
everywhere, except where they are accompanied with a maximum number of space 
and/or time derivatives of $\delta_{B{\bf q}}$ or $\delta_{\gamma{\bf q}}$, in 
which case powers of a high wave number can compensate for the smallness of 
$\chi$ or $\eta$.
Then Eqs.~(40) and (41) simplify further to
\begin{equation} 
 \frac{d}{dt}\left[\delta_{\gamma{\bf q}} -  \frac{4}{3}\delta_{B{\bf q}}\right]
=\frac{\chi T}{\rho_{\gamma}}\left[-\ddot{\delta}_{B{\bf q}}-
\frac{q^2\delta_{\gamma{\bf q}}}{4a^2}\right]\;,
\end{equation} 
\begin{eqnarray} 
&&\frac{d }{d  t}\left[a^5\left(\rho_B+\frac{4}{3}\rho_\gamma\right)\,
\Big(\dot{\delta}_{B{\bf q}}+\psi_{\bf q}\Big)\right] -\chi T 
a^3\left(\frac{q^2\dot{\delta}_\gamma}{4}+a^2\frac{d ^3\delta_{B{\bf q}}}{d  
t^3}\right)\nonumber\\
&&~~~~~~~~~~~ =-\frac{q^2a^3}{3}\rho_\gamma \delta_{\gamma{\bf q}} -
\frac{4q^2\eta a^3}{3}\dot{\delta}_{B{\bf q}}\;.~~~~~~~~
\end{eqnarray}
We also neglect terms of second order in $\chi$ and/or $\eta$, so we can set 
$\dot{\delta}_{B{\bf q}}$ equal to $3\dot{\delta}_{\gamma{\bf q}}/4$ in the 
dissipative terms in Eqs.~(42) and (43).  Then using Eq.~(42) to eliminate 
$\delta_{B{\bf q}}$ in Eq.~(43) gives our differential equation for 
$\delta_{\gamma{\bf q}}$:
\begin{eqnarray} 
&&\frac{d }{d  
t}\left[a^5\left(\rho_B+\frac{4}{3}\rho_\gamma\right)\,\Big(\frac{3}{4}
\frac{d\delta_{\gamma{\bf q}}}{dt} +\psi_{\bf q}\Big)\right] +\frac{3a^5\chi 
T\rho_B}{4\rho_\gamma}\left(\frac{3}{4}\frac{d^3{\delta}_{\gamma{\bf q}}}{d  
t^3}+ \frac{q^2}{4a^2}\frac{d\delta_{\gamma{\bf q}}}{dt} \right)\nonumber\\
&&~~~~~~~~~~~ =-\frac{q^2 a^3}{3} \rho_\gamma \delta_{\gamma{\bf q}} -\eta 
q^2a^3\frac{d\delta_{\gamma{\bf q}}}{dt}\;.~~~~~~~~
\end{eqnarray}
It will be convenient to multiply with $t^{4/3}/a^5\rho_\gamma$.  Recalling that 
$ \rho_\gamma\propto a^{-4}$, $\rho_B\propto a^{-3}$ and $a\propto t^{2/3}$, 
this gives finally
\begin{eqnarray}
&& t^{2/3}\frac{d }{d  t}\left[(1+R)t^{2/3}\frac{d \delta_{\gamma{\bf q}}}{d  
t}\right]+\frac{k^2t_L^{4/3}}{3}\delta_{\gamma{\bf q}}+\frac{\eta 
k^2t_L^{4/3}}{\rho_\gamma}
\frac{d\delta_{\gamma{\bf q}}}{dt}\nonumber\\&&~~~~~~~~~~~~~~~~+\frac{R\chi 
T}{4\rho_\gamma}\left(3t^{4/3}
\frac{d ^3\delta_{\gamma{\bf q}}}{d  t^3}+k^2t^{4/3} \frac{d\delta_{\gamma{\bf 
q}}}{dt}\right)
\nonumber\\&&~~~~=-\frac{4t^{2/3}}{3}\frac{d }{d  t}\Big((1+R)\,t^{2/3}\psi_{\bf 
q}\Big)=\frac{8N_{\bf q}}{27}(1+3R)\;,
\end{eqnarray}
where (as before) $t_L$ is some typical time of last scattering, $k\equiv q 
/a(t_L)= qt^{2/3}/t^{2/3}_La(t)$, and 
\begin{equation}
R\equiv 3\rho_B/4\rho_\gamma\propto a\;.
\end{equation}

We next turn to two different ranges of wave number in which it is possible to 
find an analytic solution of this equation.

\begin{center}
{\bf E. Solution for large $k$}
\end{center}

We consider first wave numbers that are large enough to allow the use of the WKB 
approximation.  For this purpose, we introduce a new variable 
\begin{equation}
\zeta\equiv (t/t_L)^{1/3}\;.
\end{equation} 
(This is the usual conformal time $\eta$, but with a different normalization.)  
Multiplying Eq.~(45) with $9t_L^{2/3}$ then gives
\begin{eqnarray} 
&&\frac{d}{d\zeta}\left[(1+\xi\zeta^2)\frac{d\delta_{\gamma{\bf 
q}}}{d\zeta}\right]
+3k^2t_L^2\delta_{\gamma{\bf q}}
+\frac{3\eta k^2 t_L}{\rho_\gamma\zeta^2}\frac{d\delta_{\gamma{\bf q}}}{d\zeta}
\nonumber\\&&+\frac{\xi\chi 
T}{4\rho_\gamma}\left(\frac{1}{t_L}\frac{d^3\delta_{\gamma{\bf q}}}{d\zeta^3}+
3k^2t_L\frac{d\delta_{\gamma{\bf q}}}{d\zeta}\right) =\frac{8N_{\bf 
q}t_L^{2/3}}{3}
\Big(1+3\xi\zeta^2\Big)\;.
\end{eqnarray} 
Here we have again assumed that dissipative terms are negligible except where a 
maximum number of derivatives (i.e., factors of $k$ and/or $\zeta$-derivatives)
acts on $\delta_{\gamma{\bf q}}$.  We have also used the fact that $R\propto a$ 
to set $R=\xi\zeta^2$, where $\xi$ is the ratio (46) at time $t_L$.

In the absence of dissipation, Eq.~(48) would have the  exact solution
\begin{equation}
\delta_{\gamma{\bf q}}=\frac{8N_{\bf 
q}t_L^{2/3}(1+3\xi\zeta^2)}{9(k^2t_L^2+2\xi)}\;.
\end{equation} 
(This is actually independent of our choice of $t_L$, because $\xi$ and 
$k^2t^2_L$ both scale as $t_L^{2/3}$.)  The neglect of dissipation is justified 
in this solution, because the rate of change of this expression does not yield a 
factor of the large wave number $k$ that could compensate for the smallness of 
$\chi$ and $\eta$.

To this particular solution, we must add a suitable solution of the 
corresponding homogeneous equation.   In the absence of damping we can find 
exact solutions of the form $P_\nu(i\sqrt{\xi}\zeta)$, where $P_\nu(z)$ is the 
usual Legendre function, and  $\nu$ is either of the roots of the quadratic 
equation $\nu(\nu+1)=-3k^2t_L^2/\xi$.  But this will not be useful in 
calculating the $C_\ell$ in our companion paper [1].  To get a more useful 
result, we must use the WKB approximation.

Under the assumption that
\begin{equation}
kt_L\geq \xi
\end{equation}
we can find a pair of  approximate solutions of the homogeneous equation
\begin{equation}
\delta_{\gamma{\bf q}}\propto \exp(\pm i\varphi)
\end{equation} 
with
\begin{equation}
\varphi=\sqrt{3}kt_L \int_0^\zeta \frac{d\zeta}{\sqrt{1+\xi\zeta^2}}
\end{equation}
provided we neglect dissipative terms.  Note that if $\xi$ as well as $\eta$ and 
$\chi$ were zero, then these homogeneous solutions would be exact.  More 
generally, inspection of Eq.~(48) shows that these are approximate solutions if 
the fractional rate of change of $1+\xi\zeta^2$ is small  compared with the rate 
of change of the phase $\varphi$:
$$
\frac{2\xi\zeta}{1+\xi\zeta^2}\leq \frac{ kt_L\sqrt{3}}{\sqrt{1+\xi\zeta^2}}\;.
$$
which is true at all times if and only if it is satisfied at $\zeta=1$, i.e.,
$$
kt_L\geq \frac{2\xi}{\sqrt{3(1+\xi)}}\;.
$$
For plausible values of $\xi$ this condition is actually somewhat weaker than 
Eq.~(50), but we will need the greater strength of Eq.~(50) later, when we 
calculate  the plasma velocity potential.

We can do better than Eq.~(51), and include the effects of viscosity and heat 
conduction, by seeking solutions of the form $\delta_{\gamma{\bf q}}=
A\exp(\pm i\varphi)$, with $A$ a slowly varying real amplitude.  By calculating 
the rate of change of the Wronskian of these two solutions (and replacing 
$d^3\delta_{\gamma{\bf q}}/d\zeta^3$ with $-3k^3t_L^2(1+\xi\zeta^2)^{-
1}d\delta_{\gamma{\bf q}}/d\zeta$ in the dissipative term), we easily find the 
WKB solutions of the homogenous equation:
\begin{equation} 
\delta_{\gamma{\bf q}}\propto (1+\xi\zeta^2)^{-1/4}\exp\left[\pm i\varphi-
k^2{\cal D}^2 \right]\;,
\end{equation} 
where
\begin{equation}
{\cal D}^2 =3t_L\int_0^\zeta\;\left[\frac{\eta}{2\rho_\gamma(1+\xi\zeta^2)}+
\frac{\chi T 
\xi^2\zeta^4}{8\rho_\gamma(1+\xi\zeta^2)^2}\right]\; \zeta^2\,d\zeta
\end{equation} 
The viscosity and heat conduction coefficients are given by[11]
\begin{equation}
\eta=\frac{16}{45}\rho_\gamma \tau_\gamma\;,~~~~~~~~~~\chi 
T=\frac{4}{3}\rho_\gamma\tau_\gamma\;.
\end{equation}
Here $\tau_\gamma$ is the photon mean free time
\begin{eqnarray}
\tau_\gamma &=&\frac{1}{\sigma_Tn_ec}\nonumber\\&=&
\frac{ t_L (6\pi G m_p)^{1/2}(\Omega_M/\Omega_B)^{1/2}\zeta^{9/2} 
}{\sigma_Tc(2\pi m_e k_BT_0/h^2)^{3/4}(1+z_L)^{3/4}(1-
Y/2)^{1/2}}\exp\left(\frac{\zeta^2\Delta}{2k_BT_L}\right)\;,~~~~ 
\end{eqnarray}
where $\sigma_T$ is the Thomson scattering cross section, $h$ is (only here) the 
original Planck constant, $T_0=2.738^\circ$K is the present microwave background 
temperature, $T_0=3100^\circ$K is the temperature  at last scattering, $k_B$ is 
Boltzmann's constant, $Y\simeq .23$ is the primordial helium abundance, and 
$\Delta=13.6$ eV is the hydrogen ionization energy.

The relevant solution is again the one that increases most rapidly at early 
times, 
which we can find by requiring that $\delta_{\gamma{\bf q}}\rightarrow 0$ as 
$\zeta\rightarrow 0$.  In the limit $\zeta\rightarrow 0$ the phase $\varphi$ 
vanishes as $O(\zeta)$, while ${\cal D}^2$ vanishes more rapidly because the 
mean free time of photons  is very small at early times.  Hence the linear 
combination of the particular inhomogeneous solution (49) and the homogeneous  
solutions (53) that grows most rapidly at early times  is
\begin{equation} 
\delta_{\gamma{\bf q}}=\frac{8
N_{\bf q}t_L^{2/3}}{9(k^2t_L^2+2\xi)}\Bigg[1+3\xi\zeta^2-(1+\xi\zeta^2)^{-
1/4}e^{-k^2{\cal D}^2}\cos\varphi\Bigg]\;.
\end{equation} 
(We would be able to neglect the term $2\xi$ in the denominator only under the 
condition $kt_L\geq \sqrt{2\xi}$, which for plausible values of $\xi$ is  
stronger than our assumption (50). )

To calculate the velocity potential of the plasma--photon fluid for large wave 
numbers, we will also need the rate of change of $\delta_{B{\bf q}}$.  At times 
of order $t_L$, the time derivatives of $\xi\zeta^2$, $\varphi$, and ${\cal 
D}^2$ are of the orders of $\xi/t_L$, $k$, and $\tau_L$, respectively, where 
$\tau_L$ is the photon mean free time $\tau\approx \eta/\rho_\gamma\approx \chi 
T/\rho_\gamma$ at time $t_L$.  We are assuming that $kt_L\geq \xi$, so the time 
derivative of $\varphi$ is larger than the time derivative of $\xi\zeta^2$.  
Eq.~(56) shows that damping becomes important if $k^2t_L\tau_L\geq 1$, , but 
even for such large values of $k$ we can still limit ourselves to the case
\begin{equation} 
k\tau_L\leq 1\;,
\end{equation}
in which case the time  derivative of $\varphi$ is also larger  than the time 
derivative of  $k^2{\cal D}^2$.  Hence for wave numbers $k$ in the range defined 
by Eqs.~(50) and (58),  we have
\begin{equation} 
\frac{d\delta_{\gamma{\bf q}}}{dt}\simeq \frac{8N_{\bf q}t_L^{2/3}k e^{-k^2{\cal 
D}^2}\sin\varphi }{9\sqrt{3}(1+\xi\zeta^2)^{3/4}(k^2t_L^2+2\xi)\zeta^2} \;.
\end{equation}
The dissipative terms in Eq.~(42) are smaller than this by a factor $k\tau$, so 
here we can take $\dot{\delta}_{B{\bf q}}\simeq 3\dot{\delta}_{\gamma{\bf 
q}}/4$, and Eqs.~(37), (31) and (59) then give the velocity potential
\begin{equation}
u_{\bf q}=\frac{2N_{\bf q}}{3k^2a^2(t_L)t_L^{1/3}\zeta}\left[-1+\frac{kt_Le^{-
k^2{\cal 
D}^2}\sin\varphi}{\sqrt{3}(1+\xi\zeta^2)^{3/4}(k^2t_L^2+2\xi)\zeta}\right]
\end{equation} 

\begin{center}
{\bf F. Solution for small $k$}
\end{center}

Here we can neglect viscosity and heat conduction.  For $k=0$, Eq.~(48) has an 
obvious solution
\begin{equation}
\delta_{\gamma{\bf q}}=4N_{\bf q}t_L^{2/3}\zeta^2/3=4\delta_{D{\bf q}}/3\;.
\end{equation} 
To this we can add any linear combination of the two solutions of the 
corresponding homogeneous solution, for which $\delta_{\gamma{\bf q}}$ is 
respectively 
time-independent or proportional to 
$$\int_0^\zeta \frac{d\zeta}{1+\xi\zeta^2}\;,$$
which near the beginning of the dark-matter dominated era goes as $\zeta$.  
As $\zeta$ increases these homogeneous solutions become negligible compared with 
the inhomogeneous solution (61), so at later times the solution for $k=0$ is 
given by Eq.~(61).

To get the term in $\delta_{\gamma{\bf q}}$ of first order in $k^2$, we can use 
the solution
(61) in the terms in Eq.~(48) proportional to $k^2$, so that
\begin{equation} 
\frac{d}{d\zeta}\left[\left(1+\xi\zeta^2 \right)\frac{d }{d\zeta 
}\left(\frac{3}{4}\delta_{\gamma{\bf q}} -\delta_{D{\bf q}}\right)\right]=-
3k^2t_L^2\delta_{D{\bf q}} \;.
\end{equation} 
Discarding a homogeneous term for the same reason as before, we have
\begin{equation}
\frac{d}{d\zeta}\left(\frac{3}{4}\delta_{\gamma{\bf q}}-\delta_{D{\bf 
q}}\right)=-\frac{k^2t_L^2\zeta\,\delta_{D{\bf q}}}{1+\xi\zeta^2}
\end{equation} 
which gives
\begin{equation} 
\delta_{\gamma{\bf q}}=\frac{4N_{\bf q}t_L^{2/3}}{3}\zeta^2\left[1-
\frac{k^2t_L^2}{2}\left(\frac{1}{\xi}-
\frac{1}{\zeta^2 \xi^2}\ln(1+\xi\zeta^2)\right)+\dots\right]
\end{equation} 
Also, Eqs.~(37), (27), (63) and (31)  give the plasma velocity potential for 
$k\rightarrow 0$ as
\begin{equation}
u_{\bf q}=\frac{1}{q^2}\frac{d}{dt}\left(\delta_{B{\bf q}}-\delta_{D{\bf 
q}}\right)
\rightarrow -\frac{N_{\bf q}t_L^{5/3}\zeta}{3a^2(t_L)(1+\xi\zeta^2)}\;.
\end{equation} 
As we will see, this provides a small correction to the Doppler shift, which 
for small $k$ will turn out to be mostly due to perturbations in the 
gravitational field.

\begin{center}
{\bf IV. OBSERVED TEMPERATURE FLUCTUATIONS}
\end{center}

There are three separate sources of the observed temperature fluctuation in the 
cosmic microwave background: the Sachs--Wolfe effect due to perturbations in the 
gravitational potential, the Doppler effect due to plasma peculiar velocities, 
and the intrinsic temperature fluctuations themselves.  We will consider each of 
these in turn, and then put the results together.  In calculating the 
Sachs--Wolfe and Doppler contributions, 
we will use a non-relativistic approach, taking 
the effect of the gravitational field perturbations on the observed photon 
temperature to consist entirely of the time dilation caused by a Newtonian 
gravitational potential plus the Doppler shift caused by the gravitational 
acceleration of the source and receiver.  This approach has the virtue of 
getting useful results quickly, but the results obtained in this need to be 
justified by a thoroughly relativistic 
treatment of the Sachs--Wolfe and Doppler effects, which will be given in the 
Appendix.

\begin{center}
{\bf A. Sachs--Wolfe Effect}
\end{center}

We can define a Newtonian gravitational potential $\phi$ as the solution of the 
Poisson equation
\begin{equation}
a^{-2} (t) \nabla^2\phi({\bf x},t)=4\pi G\delta\rho_{D}({\bf x}, t)
\end{equation} 
with the factor $a^{-2}$ inserted to take account of the difference between the 
Robertson--Walker co-moving coordinate vector ${\bf x}$ used here and the 
coordinate vector $a(t){\bf x}$ that measures proper distances at time $t$.
Using Eqs.~(31) and (25), this gives
\begin{eqnarray} 
\phi({\bf x},t)&=&-4\pi G\rho_D(t)t^{2/3}a^2(t)\int d^3q\;q^{-2}e^{i{\bf 
q}\cdot{\bf x}}N_{\bf q}\nonumber\\&=&
-\frac{2 a^2(t)}{3t^{4/3}}\int d^3q\;q^{-2}e^{i{\bf q}\cdot{\bf x}}N_{\bf q}
\end{eqnarray} 
It is important to note that this is time-independent during the dark matter 
era, when $a(t)\propto t^{2/3}$. 

This potential makes two separate contributions to the Sachs--Wolfe effect.  
There is a gravitational redshift, yielding a 
fractional fluctuation in the observed temperature in a direction $\hat{n}$ 
equal to $\phi(r_L\hat{n})-\phi(0)$, where $r_L$ is the Robertson--Walker radial 
coordinate of the surface of last scattering.  There is also a time-delay; if 
the unperturbed cosmic temperature reaches the value $T_L\simeq 
3000^\circ$K of last scattering at a time $t_L$, then the gravitational 
potential 
causes the cosmic temperature in a direction $\hat{n}$ to reach the value 
$T_L$ at a time $[1+\phi(r_L\hat{n})-\phi(0)]t_L$, so that the redshifted 
temperature seen now is changed by a fractional amount[12]: 
$$-\Big[t_L\dot{a}(t_L) /a(t_L)\Big] \Big[\phi(r_L\hat{n})-\phi(0)\Big]=-
\frac{2}{3}\Big[\phi(r_L\hat{n})-\phi(0)\Big]\;.$$
(This argument is valid only because $\phi$ is time-independent; otherwise we 
would have to consider the complete gravitationally delayed time-history of the 
cosmic temperature, as done in the Appendix.) 
Combining the two effects, the net fractional change in observed temperature is
\begin{equation}
\left(\frac{\Delta T({\bf n})}{T}\right)_{\rm Sachs-
Wolfe}=\frac{1}{3}\Big[\phi(r_L\hat{n})-\phi(0)\Big]\;.
\end{equation} 
As we shall see in the Appendix, this formula can be derived using the formalism 
of general relativity, which in synchronous gauge gives the famous factor of 
$1/3$ directly, without having to consider separately the gravitational red 
shift and time delay.  

It will be convenient to rewrite Eq.~(67) in terms of the physical wave number 
at the time of last scattering, $k\equiv q/a(t_L)$, so that
Eq.~(68) gives
\begin{equation}
\left(\frac{\Delta T({\bf n})}{T}\right)_{\rm Sachs-Wolfe}=
\int d^3k\; \left[e^{i{\bf k}\cdot\hat{n}d_A}-1\right]\epsilon_{\bf k}\;,
\end{equation} 
where $\epsilon_{\bf k}$ is an amplitude for fluctuations not processed by 
acoustic oscillations, defined by
\begin{equation} 
\epsilon_{\bf k}\,d^3k\equiv -\frac{2N_{\bf q}a^2(t_L)}{9q^2t_L^{4/3}}d^3q\;,
\end{equation} 
and $d_A=r_La(t_L)$ is the angular diameter distance of the surface of last 
scattering.

\begin{center}
{\bf B. Doppler Shifts}
\end{center}

The plasma velocity potential $u_{{\bf q}}$ calculated in Section III yields a 
pressure-induced plasma velocity perturbation
\begin{equation} 
{\bf v}_{\rm pressure}({\bf x},t)=a(t)\;\BM{\nabla}\int d^3q\; e^{i{\bf 
q}\cdot{\bf x}}
u_{{\bf q}}(t)\;.
\end{equation} 
(The factor $a(t)$ enters because it is the velocity in co-moving coordinates 
that is given by the co-moving gradient of the velocity potential.)
This yields a Doppler shift of the temperature of the cosmic microwave 
background seen in a direction $\hat{n}$:
\begin{eqnarray}
\left(\frac{\Delta T(\hat{n})}{T}\right)_{\rm pressure\; Doppler}&=&-i 
a(t_L)\int 
d^3q \;\hat{n}\cdot{\bf q}\,
u_{{\bf q}}(t_L) e^{i{\bf q}\cdot \hat{n}r_L}\nonumber\\&=&-i\int d^3k 
\;\hat{n}\cdot\hat{k}\,\epsilon_{\bf k}\, g(k)\, e^{i{\bf k}\cdot \hat{n}d_A}\;,
\end{eqnarray}
with the form factor $g(k)$ given by Eq.~(65) for small $k$ as
\begin{equation}
g(k)=\frac{3 k^3t_L^3}{2(1+\xi)}
\end{equation}
and for large $k$ by Eq.~(60) as
\begin{equation}
g(k)=3kt_L-\sqrt{3}(1+\xi)^{-3/4}(1+2\xi/k^2t_L^2)^{-1}e^{-
k^2d_D^2}\sin\Big(kd_H\Big)
\end{equation} 
where $\xi$ as before is $3/4$ the ratio of baryon and photon energy densities 
at the time of last scattering, ${\cal D}_L $ is the damping length ${\cal D}$ 
given in 
Eq.~(56), evaluated at $\zeta=1$ (actually, as discussed in the next section, at 
$\zeta$ a little less than unity) , and 
$d_H$ is the acoustic horizon at the time of last scattering:       
\begin{equation}
d_H= \sqrt{3}t_L\int_0^1\frac{d\zeta}{\sqrt{1+\xi\zeta^2}}=\frac{\sqrt{3}\,t_L}
{\sqrt{\xi}} \ln\left(\sqrt{\xi}+\sqrt{1+\xi}\right)\;.
\end{equation} 

In the non-relativistic approach used here, there  is also an additional 
velocity 
perturbation induced by the
gravitational potential $\phi({\bf x})$.  The proper peculiar velocity ${\bf 
v}_{\rm grav}$  produced in 
this way is given by the equation of motion[13] 
\begin{equation}
\frac{\partial}{\partial t} {\bf v}_{\rm grav}({\bf 
x},t)+\frac{\dot{a}(t)}{a(t)} {\bf v}_{\rm grav}({\bf x},t)
= -\frac{1}{a(t)}\BM{\nabla}\phi({\bf x})
\end{equation} 
Because the gravitational potential $\phi$ is time-independent, this has the 
simple solution 
\begin{equation}
{\bf v}_{\rm grav}({\bf x},t)=-a^{-1}(t)\,t\,{\BM \nabla}\phi({\bf x})=\frac{2i 
a(t)}{3\,t^{1/3}}\int d^3q\;q^{-2}{\bf q}\, e^{i{\bf q}\cdot{\bf x}}N_{\bf q}\;.
\end{equation} 
This contributes a fractional temperature shift seen in a direction $\hat{n}$:
\begin{eqnarray}
\left(\frac{\Delta T(\hat{n})}{T}\right)_{\rm gravity\;Doppler}&=&
-\hat{n}\cdot \left[{\bf v}_{\rm grav}(\hat{n}r_L,t_L)- {\bf v}_{\rm 
grav}(0,t_0)\right] \nonumber\\&=&3i\int 
d^3k\;(\hat{k}\cdot\hat{n})\,kt_L\,\epsilon_{\bf k}\left[e^{i{\bf 
k}\cdot\hat{n}d_A}-\frac{t_L^{1/3}a(t_0)}{t_0^{1/3}
a(t_L)}\right]\;,~~~~~~~
\end{eqnarray} 
where $t_0$ is the present time.
(A general relativistic derivation of this result in synchronous gauge is given 
in the Appendix.)  

\begin{center}
{\bf C. Intrinsic Temperature Fluctuations}
\end{center}

The fractional change in the photon temperature is one-fourth the fractional 
change in the photon energy  density.  The contribution of intrinsic density 
fluctuations at the time $t_L$ of last scattering to the fractional change of 
temperature seen coming from a direction $\hat{n}$ is 
therefore\fnote{\dagger}{There is a subtlety here.  To the extent that the 
opacity drops sharply from 100\% to zero, last scattering occurs at a fixed 
value of the {\em perturbed} temperature $T+\delta T$, near 3000$^\circ$K, 
rather than at a fixed value of the unperturbed temperature  or the time.   The 
effect of the intrinsic temperature fluctuation $\delta T(t)$ is thus to change 
the time of last scattering, in such a way as to produce a change $-\delta T$ in 
the value of the {\em unperturbed} temperature $T(t)$ at this time.  Since  
$T(t)\propto 1/a(t)$, we then have $\delta a/a=+\delta T/T$ at the time of last 
scattering, so that the observed temperature is shifted by the change in the 
cosmological redshift by a fractional amount $\Delta T/T=\delta a/a=+\delta 
T/T$.} 
\begin{equation}
\left(\frac{\Delta T(\hat{n})}{T}\right)_{\rm intrinsic}
=\frac{\delta\rho_\gamma({\hat n}r_L, t_L)}{4\rho_\gamma(t_L)} =\frac{1}{4}\int 
d^3q\;e^{i{\bf q}\cdot \hat{n}r_L}
\delta_{\gamma{\bf q}}(t_L)\;.
\end{equation} 
Eqs.~(57), (64) and (70) then give
\begin{equation} 
\left(\frac{\Delta T(\hat{n})}{T}\right)_{\rm intrinsic}=\int d^3k 
\;\epsilon_{\bf k}\, f(k)\, e^{i{\bf k}\cdot \hat{n}d_A}\;,
\end{equation}
with the partial form factor $f$ given by
\begin{equation}
f(k)=\left\{\begin{array}{ll}-3k^2t_L^2/2+
3\left[\xi^{-1}-\xi^{-2}\ln(1+\xi) \right] k^4t_L^4/4+\dots & k\rightarrow 0 \\
(1+2\xi/k^2t_L^2)^{-1}\left[-1-3\xi+(1+\xi)^{-1/4}e^{-
k^2{\cal D}_L^2}\cos(kd_H)\right] & 
k\;{\rm large}\;.\end{array}\right.
\end{equation} 

\begin{center}
{\bf D. Total Temperature Fluctuations}
\end{center}
We now put together the fractional temperature fluctuations given by Eqs.~(69),
(72), (78), and (80), and obtain the total fractional temperature fluctuation
\begin{equation} 
\left(\frac{\Delta T(\hat{n})}{T}\right)=\int d^3k \;\epsilon_{\bf 
k}\,\left\{\Big[ 
F(k)+i\hat{k}\cdot\hat{n}G(k)\Big]\, e^{i{\bf k}\cdot \hat{n}d_A}-1-
3i{\bf k}\cdot{\hat{n}}\frac{t_L^{4/3}a(t_0)}{t_0^{1/3}a(t_L)}\right\}\;,
\end{equation}
where $F(k)$ is the total scalar form factor, given by Eqs.~(69), (80), and (81) 
as
\begin{eqnarray} 
F(k)&=&1+f(k)\nonumber \\&=& \left\{\begin{array}{cl}
1-3k^2t_L^2/2-3\left[-\xi^{-1}+\xi^{-2}\ln(1+\xi) \right] k^4t_L^4/4+\dots & 
k\rightarrow 0 \\
(1+2\xi/k^2t_L^2)^{-1}\left[-3\xi+2\xi/k^2t_L^2+(1+\xi)^{-1/4}e^{-
k^2{\cal D}_L^2}\cos(kd_H)\right] & 
k\;{\rm large}\;.\end{array}\right. \nonumber\\&&{}
\end{eqnarray} 
and $G(k)$ is the total dipole form factor, given by Eqs.~(72), (73), (74), and 
(78) as
\begin{equation}
G(k)=3kt_L-g(k)=\left\{\begin{array}{cl}3kt_L-3k^3t_L^3/2(1+\xi)+\dots & 
k\rightarrow 0\\ \sqrt{3}(1+\xi)^{-3/4}(1+2\xi/k^2t_L^2)^{-1}e^{-
k^2{\cal D}_L^2}\sin(kd_H) & k\;{\rm large}.\end{array}\right.
\end{equation} 
The last two terms in the curly brackets in Eq.~(82) contribute only to the 
multipole coefficients $C_\ell$ for $\ell=0$ and $\ell=1$[15], and may therefore 
be dropped (as they are in Eq.~(1))  in considering the higher multipoles.

We see  that the WKB solution for large $k$ gives a poor picture of what happens 
for $k\rightarrow 0$, except in the case $\xi\ll 1$, where $d_H=\sqrt{3}t_L$, in 
which case the above pairs of expressions for $F(k)$ and $G(k)$ agree for small 
$k$.
 
As discussed in Section II, it still remains to average over the time  of last 
scattering. 
The effect of this averaging on the damping factor $\exp{-k^2{\cal D}^2}$ is 
small[15].  Otherwise, the averaging over $t$ chiefly 
affects the $\sin kd_H$ and $\cos kd_H$ factors in Eqs.~(83) and (84), which 
oscillate rapidly with the time of last scattering when $k$ is large.    
We will approximate the probability distribution of the actual time 
of last scattering $t$ as a Gaussian of the form $(1/\pi\,\Delta t)\exp(-(t-
t_L)^2/\Delta t^2)$, where $t_L$ is a nominal time of last scattering.  
Replacing $t_L$ in the sines and cosines in Eqs.~(83) and (84) with $t$, 
multiplying 
with this probability distribution, and integrating over $t$ then gives the same 
result for the form factors for large $k$ , but with an additional term now 
added to ${\cal D}_L^2$: 
\begin{equation}
\Delta{\cal D}_L^2=d_H^2\left(\frac{\Delta t}{2t_L}\right)^2\;.
\end{equation} 
This is a sort of `Landau damping,' except that the damping arises from a spread 
in the time at which the temperature of the medium is observed rather than from 
a spread in wave numbers.  
As we will see in the next section, this term makes  a smaller but not 
insignificant  contribution to the total damping.

\begin{center}
{\bf V. DISCUSSION}
\end{center}

In a companion paper[1] we show how to use the formula (82) for the total 
temperature fluctuation to derive expressions for the coefficient $C_\ell$ of 
the term of multipole number $\ell$ in the temperature fluctuation correlation 
function for general form factors $F(k)$ and $G(k)$.  As we will see there, 
the contribution of the scalar form factor $F(k)$ to $C_\ell$ arises mostly from 
wave numbers of order $\ell /d_A$ (where $d_A$ is the angular diameter distance 
of the surface of last scattering), while this approximation is much worse for 
the contribution of the dipole form factor $G(k)$.  

For the present, we will 
content ourselves with noting that if we tentatively use the WKB approximation, 
neglect damping effects, and drop the terms in  the second line of Eq.~(83) 
proportional to $\xi/k^2t_L^2$, then for $\xi$ less than $0.311$ (that is, for 
$3\xi<(1+\xi)^{-1/4}$) the squared scalar form factor $F^2(k)$ has  peaks at the 
wave numbers 
\begin{equation} 
k_n =n\pi/ d_H\;,
\end{equation}
(with $n=1, 2, \dots$), with higher peaks for odd $n$ (where the two terms in 
$F(k)$ have the same sign) than for even $n$.  The minima are at the zeroes of 
$F(k)$.  For $\xi>.311$ the only peaks are those for $n$ odd, and the minima are 
at $n$ even.  This suggests that there should be peaks  in $C_\ell$ near 
$\ell_n= (2n-1)\pi d_A/d_H$ and either lower peaks or dips near $2n\pi d_A/d_H$, 
depending on the value of $\xi$.  These peaks are known as the Doppler peaks 
(though Eq.~(84) shows that the contribution of the Doppler shift 
is very small at all the wave numbers $k_n$.)  These results depend critically 
on the negative sign of the term $-3\xi$ in the second line of Eq.~(83); if this 
term had turned out to be positive then for $\xi>.311$ the positions of the 
peaks and dips would be interchanged.  Despite what is sometimes said[16], there 
is no way without detailed calculations to see that the first Doppler peak 
should be at $\ell\simeq \pi d_A/d_H$, rather than at a multipole number twice 
as large.

We can now check whether the WKB approximation used in subsection 3E is valid at 
the first Doppler peak.  According to  Eqs.~(86) and (75), we have
\begin{equation} 
k_1t_L=\frac{\pi\sqrt{\xi}}{\sqrt{3}\ln(\sqrt{\xi}+\sqrt{1+\xi})}\;,
\end{equation} 
so the ratio of the wave number at the first Doppler peak to the mininum wave 
number $k_{\rm min}$ allowed by the inequality (50) is
\begin{equation}
\frac{k_1}{k_{\rm min}}=
\frac{\pi}{\sqrt{3\xi}\,\ln(\sqrt{\xi}+\sqrt{1+\xi})}\;.
\end{equation} 
The WKB approximation is valid at wave numbers down to the first Doppler peak if 
this ratio is sufficiently larger than unity.  For instance, for 
$\Omega_Bh^2=0.03$ we 
have $\xi=0.81$, so Eq.~(88) gives $ k_1/k_{\rm min} =2.5$, making the WKB 
approximation fairly good at the first Doppler peak.  The WKB approximation is 
somewhat better at $k_1$ for smaller values of $\Omega_Bh^2$, though it still 
breaks down at smaller wave numbers unless  $\Omega_Bh^2=0$.  For all plausible 
values of $\xi$ the WKB approximation is excellent at the higher Doppler peaks.

Next, let us consider the importance of damping.  It might seem that we should 
calculate the damping length ${\cal D}_L$ by integrating in Eq.~(54) up to the  
time of last scattering, corresponding to $\zeta=1$.  But at the nominal time of 
last scattering (defined so that the probability of any future scattering is 
50\%), the photon collision rate $1/\tau_\gamma$ given by Eq.~(56)  is $0.2 
\sqrt{\Omega_B/\Omega_M}/t_L$, which is already considerably smaller  than the 
expansion rate $2/3t_L$, so that we cannot trust the hydrodynamic calculations 
used to obtain Eq.~(54).  We will instead 
integrate in Eq.~(54) only up to a value $\zeta_{\rm max}$ of $\zeta$ at which
the photon collision rate becomes equal to the expansion rate, and set
$$
\tau_\gamma\simeq \frac{3t_L}{2} \left(\frac{\zeta}{\zeta_{\rm 
max}}\right)^{9/2}\exp\left(-\frac{\Delta}{2k_BT_L}(\zeta_{\rm max}^2-
\zeta^2)\right) \;.
$$
The exponential factor (with $\Delta/2k_BT_L=25.5$) is so sharply peaked at 
$\zeta=\zeta_{max}$ that we can approximate $\zeta_{\rm max}^2-\zeta^2\simeq 
2\zeta_{\rm max}(\zeta_{\rm max}-\zeta)$ in the exponent and set $\zeta$ equal 
to $\zeta_{\rm max}$ everywhere else in the integral, giving
$$
{\cal D}_L^2\simeq \frac{3 t_L^2}{2\zeta_{\rm 
max}^3}\left(\frac{8}{15(1+\xi\zeta_{\rm max}^2)}+\frac{\xi^2\zeta_{\rm 
max}^4}{2(1+\xi\zeta_{\rm max}^2)^2}\right)\left(
\frac{k_BT_L}{\Delta}\right) 
$$ 
Furthermore, $\zeta_{\rm max}$ is very close to unity.  (For instance, for  
$\Omega_M/\Omega_B=7.5$, we have $\zeta_{\rm max}=0.96$.  That is, we carry the 
damping integral down to a temperature $T_L/\zeta_{\rm max}^2\simeq 3360^\circ$K 
instead of 3100$^\circ$K.)  Hence in this result we may as well replace 
$\zeta_{\rm max}$ with unity, so that 
$$
{\cal D}_L^2\simeq \frac{3 
t_L^2}{2}\left(\frac{8}{15(1+\xi)}+\frac{\xi^2}{2(1+\xi)^2}\right)\left(
\frac{k_BT_L}{\Delta}\right) 
$$
This approximation leads to the additional simplification that the damping 
length is independent of most of the parameters appearing in Eq.~(56), including 
the ratio $\Omega_B/\Omega_M$.

There is a smaller additional contribution 
from the averaging over oscillatory terms, given by Eq.~(85).  
To evaluate this Landau damping term, we will need the ratio $\Delta t/t_L$. We 
noted in Section II that the probability that a photon will not be scattered 
again rises from 2\% at about 3400$^\circ$K to 98\% at about 2800$^\circ$K, with 
very little dependence on any cosmological parameters. Matching this to the 
probabilities calculated from  the approximation that the probability of 
scattering in a time interval from $t$ to $t+\Delta t$ is a Gaussian 
$(dt/\pi\,\Delta t)\exp(-(t-t_L)^2/\Delta t^2)$,  and using the relation 
$T\propto t^{-2/3}$, we find $\Delta t/t_L=0.10$, so the contribution to ${\cal 
D}_L^2$ in Eq.~(85)  has a value $0.0025d_H^2$.  Adding this to the quantity we 
have calculated from the integral (54) gives the total squared damping length
\begin{equation}
d_D^2\equiv {\cal D}_L^2+\Delta{\cal D}_L^2\simeq 
0.029\,t_L^2\left(\frac{8}{15(1+\xi)}+\frac{\xi^2}{2(1+\xi)^2}\right)+0.0025\,
d_H^2\;.
\end{equation}

For instance, for $\Omega_Bh^2=0.02$ (so that $\xi=0.54$) Eq.~(75) gives
$d_H=1.61\,t_L$, so $d_D^2=0.0071\, d_H^2$.  Hence 
at the first Doppler peak the argument of the damping exponential is $
d_D^2k_1^2\simeq 0.07$.  (This depends very little on $\xi$.)  We see that 
damping is not  important at the first Dopper peak, in agreement with more 
accurate computer calculations[17], but is quite significant at the second 
Doppler peak.  
One effect of damping is to shift the second and higher Doppler peaks to lower 
values of $k$ and $\ell$.

In deriving the wave numbers (86) of the Doppler peaks we also neglected the 
terms proportional to $\xi/k^2t_L^2$ in the second line of Eq.~(83).  At the 
first Doppler peak this quantity is given by Eq.~(75) as
$$
\frac{\xi}{k_1^2t_L^2}=\frac{3}{\pi^2}\Big[\ln(\sqrt{\xi}+\sqrt{1+\xi})\Big]^2\;
.
$$ 
This is 0.20 for $\Omega_Bh^2=0.03$, for which $\xi=0.81$, and less for smaller 
values of  $\Omega_Bh^2$.  This approximation is thus fair at the first Doppler 
peak, and becomes excellent at the higher Doppler peaks.

Finally, we must ask what values of $k$ are small enough so that we can ignore 
acoustic oscillations during the era when the photon energy density exceeded the 
dark matter plus baryon density, during which our analysis does not apply.  
During this era the Robertson--Walker scale factor $a(t)$ went as $t^{1/2}$, and 
the speed of sound was $1/\sqrt{3}$, so the phase change of  acoustic 
oscillations up to the time $t_C$ of the crossover from radiation dominance to 
matter dominance was
$$
\Delta\varphi=q\int^{t_C}_0 
\frac{dt}{\sqrt{3}a(t)}=\frac{2qt_C}{\sqrt{3}a(t_C)}= 
\frac{2kt_L}{\sqrt{3}}\left(\frac{t_C\,a(t_L)}{t_L\,a(t_C)}\right)\;.
$$
The redshift $z_C$ at the crossover is given by 
$1+z_C=\Omega_M/\Omega_\gamma=4\times 10^4 \Omega_Mh^2$.  During the period from 
this crossover to the present
the scale factor $a(t)$ went as $t^{2/3}$, so the ratio in parentheses is
$$ 
\frac{t_C\,a(t_L)}{t_L\,a(t_C)}=\sqrt{\frac{1+z_L}{1+z_C}}=\frac{1}{6.0\,
\sqrt{\Omega_Mh^2}}\;.
$$
Using this and Eq.~(75) gives
\begin{equation}
\Delta\varphi\simeq \frac{0.35}{\sqrt{\Omega_M h^2}}\,\left(\frac{k}{k_1}\right) 
\frac{\sqrt{\xi}}{\ln(\sqrt{\xi}+\sqrt{1+\xi})}\;.
\end{equation} 
For instance, if we take $\Omega_Mh^2=0.15$ and $\Omega_Bh^2=0.03$, then 
$\Delta\varphi\simeq 1 $ at the first Doppler peak, indicating that oscillations 
in the radiation-dominated era are becoming important at the first Doppler peak.  
This is not to say that we are making an error of order unity in the argument 
$\varphi$ of the sines and cosines in Eqs.~(83) and (84), but rather that the 
evolution of the perturbations during this much of their oscillations has not 
been reliably calculated.  This source of error is mitigated in reference [1] by 
including the effects of photon and neutrino energies on $a(t)$ in calculating 
the horizon distance.

Our formula (84) for the dipole form factor $G(k)$ raises the possibility of a 
maximum in $G(k)$ at $kd_H=\pi/2$, yielding a ``zeroth Doppler peak,''  produced 
(as the first Doppler peak is not) by the Doppler effect.  For 
$\Omega_Bh^2=0.03$ the wave number at this supposed peak is too small for us to 
trust the WKB approximation used to derive Eq.~(84) at this peak, but the peak 
in $G(k)$ at $kd_H=\pi/2$ would  definitely be there for much smaller values of 
$\Omega_Bh^2$.  In particular, the calculations of reference [1] show such a 
zeroth Doppler peak in $C_\ell$ at $\ell\simeq 0.45d_A/d_H$ for $\Omega_B=0$.

\begin{center}
{\bf APPENDIX: RELATIVISTIC CALCULATION OF THE SACHS--WOLFE AND DOPPLER EFFECTS}
\end{center}

In Section IV we gave a  derivation of the Sachs--Wolfe and Doppler effects, 
using heuristic arguments to supplement relativistic results.  For completeness, 
this Appendix will present a thouroughly relativistic 
derivation in synchronous gauge, taking into account the possible presence of a 
vacuum energy, which may or may not be constant.  This goes over familiar 
ground, first considered by 
Sachs and Wolfe[18], but I as far as I know there is no published treatment of 
the `integrated Sachs--Wolfe effect' in synchronous gauge that goes explicitly 
and analytically into 
the details presented here, including the possibility of a varying vacuum 
energy.

A light ray travelling toward the center of the 
Robertson--Walker coordinate system from the direction $\hat{n}$ 
will have a co-moving radial coordinate $r$ related to $t$ by 
\begin{equation} 
0= {g}_{\mu\nu\;{\rm total}}dx^\mu dx^\nu=-dt^2+\left(a^2(t) 
+h_{rr}(r\hat{n},t)\right)dr^2\;,
\end{equation} 
or in other words
\begin{equation} 
\frac{dr}{dt}=-\left(a^2+h_{rr}\right)^{-1/2}\simeq -
\frac{1}{a}+\frac{h_{rr}}{2a^3}\;.
\end{equation} 
The first-order solution is
\begin{equation} 
r(t)=s(t)+\frac{1}{2}\int_{t_L}^t\frac{dt'}{a^3(t')}h_{rr}\left(s(t')\hat{n},t'
\right)\;,
\end{equation} 
where $s(t)$ is the zero-th order solution for the radial coordinate which has 
the value $r_L$ at $t=t_L$:
\begin{equation} 
s(t)=r_L-\int_{t_L}^t\frac{dt'}{a(t')}\;.
\end{equation} 
In particular, if the ray reaches $r=0$ at a time $t_0$,  then
\begin{equation} 
0=s(t_0) +\frac{1}{2}\int_{t_L}^{t_0}\frac{dt}{a^3(t)}h_{rr}\left(s(t) 
\hat{n},t\right)\;.
\end{equation}

A time interval $\delta t_L$ between successive light wave crests at the time 
$t_L$ of last scattering produces a time interval $\delta t_0$ at $t_0$ given by 
the variation of Eq.~(95):
\begin{eqnarray} 
&&0=\delta t_L\left[\frac{1}{a(t_L)}-\frac{1}{2}\frac{h_{rr}(r_L 
\hat{n},t_L)}{a^3(t_L)}+\frac{1}{2\,a(t_L)}\int_{t_L}^{t_0}\frac{dt}{a^3(t)}
\left(\frac{\partial h_{rr}(r\hat{n},t)}{\partial r}\right)_{r=s(t)}\right]
\nonumber\\&&+ \delta t_L\left(\frac{\partial u(r \hat{n},t_L)}{\partial 
r}\right)_{r=r_L}+\delta t_0\left[ -
\frac{1}{a(t_0)}+\frac{1}{2}\frac{h_{rr}(0,t_0)}{a^3(t_0)}\right] 
\;.
\end{eqnarray} 
(The velocity potential term on the right-hand side arises from the pressure-
induced change with time of the radial coordinate $r_L$ of the light source in 
Eq.~(95).)
The total rate of change of the quantity $h_{rr}(s(t) \hat{n},t)/a^2(t)$ in 
Eq.~(96) is 
$$\frac{d}{dt}\frac{h_{rr}(s(t) \hat{n},t)}{a^2(t)} = 
\left(\frac{\partial}{\partial 
t}\frac{h_{rr}(r\hat{n},t)}{a^2(t)}\right)_{r=s(t)}-
\frac{1}{a^3(t)}\left(\frac{\partial h_{rr}(r\hat{n},t)}{\partial 
r}\right)_{r=s(t)}\;, $$
so Eq.~(96) may be written
\begin{eqnarray}
&&0=\delta t_L\left[\frac{1}{a(t_L)}-
\frac{1}{2}\frac{h_{rr}(0,t_0)}{a^2(t_0)a(t_L)}+\frac{1}{2\,a(t_L)}\int_{t_L}^{t
_0} dt\,\left\{\frac{\partial }{\partial t}
\left(\frac{h_{rr}(r\hat{n},t)}{a^2(t)}\right)\right\}_{r=s(t)}\right] 
\nonumber\\&&
+\delta t_L\left(\frac{\partial u(r \hat{n},t)}{\partial r}\right)_{r=r_L}
+\delta t_0\left[ -
\frac{1}{a(t_0)}+\frac{1}{2}\frac{h_{rr}(0,t_0)}{a^3(t_0)}\right]\;.
\end{eqnarray} 
Hence to first order the ratio of the received and emitted frequencies is
\begin{eqnarray}
\frac{\nu_0}{\nu_L}&=&\frac{\delta t_L}{\delta 
t_0}=\frac{a(t_L)}{a(t_0)}\left[1-\frac{1}{2}\int_{t_0}^{t_L}
\left\{\frac{\partial}{\partial t}\left(\frac{h_{rr}(r\hat{n},t)}{a^2(t)}
\right)\right\}_{r=s(t)}\right.
\nonumber\\&&-\left.a(t_L)\left(\frac{\partial u(r\hat{n},t)}{\partial 
r}\right)_{r=r_L}\right]\;.
\end{eqnarray}
This gives a fractional shift in the radiation temperature observed at time 
$t_0$ coming from direction $\hat{n}$, from its unperturbed value: 
$T_0=T_La(t_L)/a(t_0)$: 
\begin{eqnarray}
&&\left(\frac{\Delta T(\hat{n})}{T}\right)_{\rm SW,\, 
Dop}=\frac{\nu_0}{a(t_L)\nu_L/a(t_0)}-1\nonumber\\&=&-\int_{t_L}^{t_0} 
dt\,\left\{\frac{\partial }{\partial t}
\left(\frac{h_{rr}(r\hat{n},t)}{2\,a^2(t)}\right)\right\}_{r=s(t)} - 
a(t_L)\left(\frac{\partial u(r_L \hat{n},t_L)}{\partial r}\right)_{r=r_L}\,.~~~ 
\end{eqnarray} 

Now we have to think about how to relate the $rr$ component of the metric 
perturbation to the field $\psi$ appearing in Section III.  In 
general, the metric perturbation may be written as
\begin{equation}
h_{ij}=A\delta_{ij}+\frac{\partial^2 B}{\partial x^i\partial x^j}\;.
\end{equation} 
The quantity entering into the integrand in Eq.~(99) is then
\begin{equation} 
\frac{\partial }{\partial t}
\left(\frac{h_{rr}(r\hat{n},t)}{a^2(t)}\right)=\alpha(r\hat{n},t) 
+\frac{\partial^2\beta(r\hat{n},t)}{\partial r^2}\;,
\end{equation} 
where
\begin{equation} 
\alpha\equiv \frac{\partial}{\partial 
t}\left(\frac{A}{2a^2}\right)\;,\qquad\qquad
\beta\equiv \frac{\partial}{\partial t}\left(\frac{B}{2a^2}\right)\;.
\end{equation} 
The field $\psi$ defined by Eq.~(23) is given by
\begin{equation}
\psi=3\,\alpha+\nabla^2\beta\;.
\end{equation} 
We also need a relation between $\alpha$ and $\beta$, which can be taken from 
the field equation for the full metric perturbation[19]:
\begin{eqnarray}
&&\nabla^2h_{ij}-\frac{\partial^2 h_{ik}}{\partial x^j \partial x^k}
-\frac{\partial^2 h_{jk}}{\partial x^i \partial x^k}+\frac{\partial^2 
h_{kk}}{\partial x^i \partial x^j}\nonumber\\
&&-a^2\ddot{h}_{ij}+a\dot{a}\left(\dot{h}_{ij}-\delta_{ij}\dot{h}_{kk}\right)
+2\dot{a}^2\delta_{ij}h_{kk}+2a\ddot{a}h_{ij}\nonumber\\
&&=-8\pi G\,\left(\delta\varrho-\delta p\right)a^4\delta_{ij}\;.
\end{eqnarray}
(For simplicity we are here taking the universe to be spatially flat, which is 
certainly a good approximation at high redshifts, and seems to be a good 
approximation even at present.)
The $\partial^2/\partial x^i\partial x^j$ terms in Eq.~(104) give
\begin{equation} 
A=a^2\ddot{B}-a\dot{a}\dot{B}-2a\ddot{a}B=a\frac{\partial}{\partial 
t}\left(a^3\frac{\partial}{\partial t}\left(a^{-2}B\right)\right)\;.
\end{equation} 
In terms of the quantities defined by Eq.~(102), this is
\begin{equation}
\alpha=\frac{\partial}{\partial t}\left(\frac{1}{a}\frac{\partial}{\partial 
t}\left(a^3\beta\right)\right)\;.
\end{equation} 
Hence for a given gravitational potential $\psi$, we can calculate $\beta$ 
by solving Eq.~(103):
\begin{equation}
\frac{\partial}{\partial t}\left(\frac{1}{a}\frac{\partial}{\partial 
t}\left(a^3\beta\right)\right)+\nabla^2\beta=\psi
\end{equation} 
and then use Eq.~(106) to find $\alpha$.

Now we return to the fractional temperature shift (99).  Using Eqs.~(101) and 
(106) lets us write this as
\begin{eqnarray} 
\left(\frac{\Delta T(\hat{n})}{T}\right)_{\rm SW,\, 
Dop}&=&-\int_{t_L}^{t_0} 
dt\,\left(\frac{\partial^2\beta(r\hat{n},t)}{\partial r^2}\right)_{r=s(t)} - 
a(t_L)\left(\frac{\partial u(r_L \hat{n},t_L)}{\partial 
r}\right)_{r=r_L}\nonumber\\&&-\int_{t_L}^{t_0}\left(\frac{\partial}{\partial 
t}\left(\frac{1}{a}\frac{\partial}{\partial 
t}(a^3(t)\,\beta(r\hat{n},t))\right)\right)_{r=s(t)}\,.~~~ 
\end{eqnarray} 
 To do the first integral here 
we note that
\begin{eqnarray} 
&&\left(\frac{\partial^2\beta(r\hat{n},t)}{\partial r^2}\right)_{r=s(t)}
\nonumber\\&&~~~~~~~=-\frac{d}{dt}\left[\left(a^2(t) 
\frac{\partial\beta(r\hat{n},t)}{\partial t} 
+a(t)\dot{a}(t)\beta(r\hat{n},t)+a(t)\frac{\partial\beta(r\hat{n},t)}{\partial 
r}\right)_{r=s(t)}\right]
\nonumber\\
&&+\left(a^2(t)\frac{\partial^2\beta(r\hat{n},t)}{\partial t^2}
+3a(t)\dot{a}(t) \frac{\partial\beta(r\hat{n},t)}{\partial t}+
\Big(a(t)\ddot{a}(t)+\dot{a}^2(t)\Big)\, \beta(r\hat{n},t)\right)_{r=s(t)}~~~~~
\nonumber\\{}
\end{eqnarray}
The fractional temperature fluctuation (108) may therefore be written
\begin{equation}
\left(\frac{\Delta T(\hat{n})}{T}\right)_{\rm SW,\, 
Dop}=\left(\frac{\Delta T(\hat{n})}{T}\right)_{\rm early }+\left(\frac{\Delta 
T(\hat{n})}{T}\right)_{\rm late }+
\left(\frac{\Delta T(\hat{n})}{T}\right)_{\rm integrated}\;,
\end{equation} 
where
\begin{eqnarray}
&&\left(\frac{\Delta T(\hat{n})}{T}\right)_{\rm early }=-a^2(t_L) 
\left(\frac{\partial\beta(r_L\hat{n},t)}{\partial t}\right)_{t=t_L}-
a(t_L)\dot{a}(t_L)\beta(r_L\hat{n},t_L)
\nonumber\\
&&~~~~~~~~~ -a(t_L)\left(\frac{\partial
\beta(r\hat{n},t_L)}{\partial r}\right)_{r=r_L}- a(t_L)\left(\frac{\partial 
u(r_L \hat{n},t_L)}{\partial r}\right)_{r=r_L}\\
&&\left(\frac{\Delta T(\hat{n})}{T}\right)_{\rm late }= a^2(t_0) 
\left(\frac{\partial\beta(0,t)}{\partial t}\right)_{t=t_0}
+a(t_0)\dot{a}(t_0)\beta(0,t_0)\nonumber\\
&&
~~~~~~~~~ +a(t_0)\left(\frac{\partial
\beta(r\hat{n},t_0)}{\partial r}\right)_{r=0}\\&&\left(\frac{\Delta 
T(\hat{n})}{T}\right)_{\rm integrated }=
\nonumber\\&&-2\int_{t_0}^{t_L} 
dt\,\left(a^2(t)\frac{\partial^2\beta(r\hat{n},t)}{\partial t^2}
+4a(t)\dot{a}(t) \frac{\partial\beta(r\hat{n},t)}{\partial t}+
2\Big(a(t)\ddot{a}(t)+\dot{a}^2(t)\Big)\, 
\beta(r\hat{n},t)\right)_{r=s(t)}\;.\nonumber\\&&{}
\end{eqnarray}

In evaluating these three contributions to the temperature fluctuation, it is
helpful to note a relation between $\beta$ and the conventionally defined 
Newtonian potential $\phi$ that applies not only for a gravitational field 
dominated by cold dark matter, but also in the presence of a constant vacuum 
energy. 
Combining Eqs.~(26) and (27) gives
\begin{equation}
\frac{\partial }{\partial t}\left(\frac{1}{4\pi G a^2\rho_D}\frac{\partial 
}{\partial t}a^2\psi\right)
=\psi\;.
\end{equation} 
Taking into account the relation $\rho_D\propto a^{-3}$, an elementary 
manipulation then gives
\begin{equation}
\frac{\partial}{\partial t}\left(\frac{1}{a}\frac{\partial }{\partial 
t}a^3\psi\right)
=\left[4\pi G\rho_D+\frac{d }{d t}\left(\frac{\dot{a}}{a}\right)
\right]a^2\psi
\end{equation} 
The equations of the Friedmann model give 
\begin{equation}
\frac{d}{dt}\left(\frac{\dot{a}}{a}\right)=-4\pi G(\rho+p)\;.
\end{equation} 
A constant vacuum energy density $\rho_V$ is associated with a 
pressure $p_V=-\rho_V$, while cold dark matter by definition has
zero pressure, so as long as the gravitational field is dominated by cold dark 
matter and a constant vacuum energy, the right-hand side of Eq.~(116) is $-4\pi 
G \rho_V$, and Eq.~(115) then gives 
\begin{equation} 
\frac{\partial }{\partial t}\left(\frac{1}{a}\frac{\partial }{\partial 
t}a^3\psi\right)=0\;.
\end{equation} 
Comparing Eq.~(106) with Eq.~(110), we now see that Eqs.~(107) has 
the solution
\begin{equation}
\nabla^2\beta=\psi
\end{equation} 
More specifically, if we define a Newtonian gravitational potential $\phi$ by
Poisson's equation
\begin{equation}
a^{-2}\nabla^2\phi=4\pi G\delta\varrho_{D}\;,
\end{equation} 
then Eqs.~(26) and (118)  show that the Newtonian potential is 
\begin{equation}
\phi=-\frac{\partial}{\partial t}\Big(a^2\beta\Big)\;.
\end{equation} 

This result is not applicable if the gravitational field receives significant 
cotnributions from a varying vacuum energy, but 
even in quintessence theories it is reasonable to assume that a  vacuum energy 
density of any sort is negligible at and near the time of last scattering.  
(It certainly must be much less than the radiation energy density at the time of 
cosmological nucleosynthesis, in order to avoid the production of too much 
helium.)  We have also been relying here on the approximation that the radiation 
energy density is much less than the dark matter density at around the time of 
last scattering.    
Therefore  the early-time contribution (111) to the temperature 
fluctuation can be calculated using the relation (118) and $\psi\propto t^{-
1/3}$, which give $\beta \propto t^{-1/3}$.  Since here
$a\propto t^{2/3}$, Eq.~(120) then gives $\beta=-t\phi/a^2$, with 
$\phi$ time-independent.   The early-time contribution (111) to the temperature 
fluctuation may therefore be expressed as
\begin{equation}
\left(\frac{\Delta T(\hat{n})}{T}\right)_{\rm early }=\frac{1}{3}\phi(r\hat{n}) 
+ \frac{t_L}{a(t_L)}\left(\frac{\partial
\phi(r\hat{n})}{\partial r}\right)_{r=r_L}- a(t_L)\left(\frac{\partial u(r_L 
\hat{n},t_L)}{\partial r}\right)_{r=r_L}\;.
\end{equation} 
This yields the Sachs--Wolfe temperature shift (68) and the gravitationally-
induced Doppler shift (77) (aside from the terms arising from $r=0$, about which 
more later), as well as the pressure-induced Doppler shift (72).  The famous 
factor $1/3$ in the first term 
on the right-hand side arises in `Newtonian gauge' as the sum of a gravitational 
redshift equal to $\phi$, and a term in the intrinsic temperature fluctuation 
equal to $-2\phi/3$, while in the synchronous gauge used here this term is due 
entirely to the metric perturbation.  It is a curious feature of synchronous 
gauge that what we have called the gravitationally-induced Doppler shift also 
arises from the metric perturbation.

It is not appropriate to neglect the 
vacuum energy  at $t=t_0$, so it cannot be ignored in the early-time 
contribution (112) to the temperature fluctuation. Therefore in general this 
contribution is {\em not} the same as the $r=0$ terms in Eqs.~(68) and (77).  
Nevertheless, the terms in the early-time contribution to the temperature 
fluctuation are only of zeroth and first 
order in $\hat{n}$ (like the $r=0$ terms in Eqs.~(68) and (77)) so these terms 
can only affect the multipole coefficients for $\ell=0$ and $\ell=1$.  

This leaves the integrated term (113) as the only correction to 
the results of Section IV for $\ell\geq 2$.  The integrand vanishes if we ignore 
the vacuum energy and radiation energy, in which case $a\propto t^{2/3}$ and 
$\beta\propto 
t^{-1/3}$, so the integral receives a contribution only for $t$ near $t_0$, and 
is therefore expected to be a small correction[20].  Furthermore, although this 
integral is fairly complicated, 
it has a simple dependence on $\hat{n}$.  In the presence of a vacuum energy, 
$\psi({\bf x},t)$ can have a fairly complicated dependence on time, but, without 
pressure forces acting on the dark matter, its ${\bf x}$ dependence is the same 
as we found in the absence of vacuum energy, given by Eqs.~(31) and (70) as
\begin{equation}
\psi({\bf x},t)=f(t)\,\int d^3k\;e^{i{\bf k}\cdot {\bf x}}{\bf k}^2\epsilon_{\bf 
k}
\end{equation} 
with $f(t)$ not proportional to $t^{-1/3}$ where the vacuum energy is 
appreciable.  For a constant vacuum energy, $\beta$ is then given by Eq.~(118) 
as
\begin{equation}
\beta({\bf x},t)=-f(t)\,\int d^3k\;e^{i{\bf k}\cdot {\bf x}}\epsilon_{\bf k}
\end{equation} 
The ``integrated'' contribution (113) to the temperature fluctuation then 
takes the form
\begin{eqnarray} 
&&\left(\frac{\Delta T(\hat{n})}{T}\right)_{\rm integrated }=2\int 
d^3k\;\int_{t_0}^{t_L} 
dt\,e^{i{\bf k}\cdot \hat{n}s(t)}\epsilon_{\bf k}
\nonumber\\&&\times \,\left(a^2(t)\ddot{f}(t)
+4a(t)\dot{a}(t)\dot{f}(t)+
2\Big(a(t)\ddot{a}(t)+\dot{a}^2(t)\Big)f(t)\right) \;.
\end{eqnarray} 
It can be shown that this makes an additive contribution to $\ell(\ell+1)C_\ell$ 
that for large $\ell$ goes as $1/\ell$, with no interference between
this contribution to the temperature fluctuation and the other 
contributions.[21]  For a time-varying (but spatially constant) vacuum energy 
the function $\beta({\bf x},t)$ does not satisfy the relations (118) and (123),
but Eq.~(107) shows that its spatial Fourier transform is nevertheless just 
proportional to $\epsilon_{\bf k}$ for large $k$, so the integrated term
still makes a contribution to $\ell(\ell+1)C_\ell$ that is proportional to
$1/\ell$ for large $\ell$.

\begin{center}
{\bf ACKNOWLEDGMENTS}
\end{center}

I am grateful for helpful correspondence with E. Bertschinger, J. R. Bond, L. P. 
Grishchuk, and 
M. White. This 
research was supported in part by the Robert A. Welch Foundation and NSF Grants 
PHY-0071512 and PHY-9511632.

\begin{center}
{\bf REFERENCES}
\end{center}

\begin{enumerate}
\item S. Weinberg, ``Fluctuations in the Cosmic Microwave Background II: 
$C_\ell$ for large and small $\ell$,'' UTTG-04-01, astro-ph/0103281. 

\item  E. R. Harrison, Phys. Rev. {\bf D1}, 2726 (1970);
P. J. E. Peebles and J. T. Yu,  Ap. J. {\bf 162}, 815 (1970);
Ya. B. Zel'dovich, Astron. Ap. {\bf 5}, 84 (1970).

\item S. Hawking,  Phys. Lett. {\bf 115B}, 295 (1982);  A. A. Starobinsky, 
 Phys. Lett.  117B, 175 (1982); A. Guth and S.-Y. Pi, Phys. Rev. 
Lett. {\bf 49}, 1110 (1982); J. M. Bardeen, P. J. Steinhardt, and M. S. Turner, 
 Phys. Rev. {\bf D28}, 679 (1983); W. Fischler, B. Ratra, and L. Susskind, 
 Nucl. Phys. {\bf B259}, 730 (1985).

\item W. Hu, U. Seljak, M. White, and M. Zaldarriaga,  Phys. Rev. {\bf D57}, 
3290 (1998),  and earlier references cited therein.  For analysis of recent 
observations, see J. R. Bond  et al., astro-ph/0011378 (2000).

\item P. J. E. Peebles and J. T. Yu, ref. 2;
J. R. Bond and G. Efstathiou,  Ap. J. Lett. {\bf 285}, L45 (1984);  Mon. Not. 
Roy. Astron. Soc. {\bf 226}, 655 (1987); A. G. Doroshkevich, Sov. Astron. Lett. 
{\bf 14}, 125 (1988); F. Atrio-Barandela, A. G. Doroshkevich, and A. A. Klypin, 
Astrophys. J. {\bf 378}, 1 (1991); P. Naselsky and I. 
Novikov, Ap. J. {\bf 413}, 14 (1993); H. E. 
J$\phi$rgensen, E. Kotok, P. Naselsky, and I. Novikov,  Astron. Astrophys. {\bf 
294}, 639 (1995); C-P. Ma and E. Bertschinger, Ap. J. 
{\bf 455}, 7 (1995).  For comments on some of these articles, see reference [6].  
The most up-to-date and comprehensive calculation is that of  W. Hu and N. 
Sugiyama,  Ap. J. {\bf 444}, 489 (1995); {\bf 
471}, 542 (1996), but they do not collect their results into a single formula 
for the temperature shift, so it is not easy to compare their results with those 
of the present paper.

\item Several authors have given approximate formulas for the temperature 
fluctuation that fit the general form of Eq.~(1).  In particular, the results 
given in Eqs.~(8) and (9) can be obtained from Eq.~(1) of Naselsky and 
Novikov, ref. 5, by applying some corrections:  the 
factor $(1+\nu)$ should be omitted in their definition of $\xi$ (in their 
notation, $\epsilon$); the factor $(1+z)^{-1}$ should be omitted in their 
definition of their parameter $\omega$, so that their $\omega$ equals $\xi$; and 
a factor $\omega$ should be included in the last term of the numerator in the 
argument of the logarithm of their Eq.~(2); and the damping factor $\exp(-
k^2d_D^2)$ and terms proportional to 
$\xi/k^2t_L^2$ should be inserted.  This paper did not give the derivation of 
their Eq.~(1), but quoted Doroshkevich, ref. 5, despite the fact that their 
result was quite different from that of Doroshkevich.  There was an obvious 
misprint in the formula given by Doroshkevich, but this was corrected in the 
later paper by Atrio-Barandela, Doroshkevich, and Klypin, ref. 5.  The formula 
in this paper includes the terms proportional to   $\xi/k^2t_L^2$, which had 
been omitted by Naselsky and Novikov, but omitted the  factors of $(1+\xi)$ that 
had been included by Naselsky and Novikov, and also included a spurious term in 
the analog of $G(k)$ (the 1 in the numerator of the second line of their 
Eq.~(4)).  This paper gave differential equations for the time development of 
the perturbations, but did not explain how they were used to calculate the 
temperature fluctuation.  A few years  later the  formula given by Naselsky and 
Novikov was repeated by 
J$\phi$rgensen, Kotok, Naselsky, and Novikov, ref. 5, and the differential 
equations on which the formula was based 
were given.  However,  once again the derivation of the formula from these 
equations was 
not explained, and an overly restrictive lower bound on $k$ was given for the 
validity of the formula, that $k$ must be larger than the inverse conformal 
time.  If this condition were really necessary, then the formula would not be 
applicable at the first Doppler peak.   We will see in Section III that the 
lower bound on $k$ is actually  less restrictive, and in particular disappears 
for $\xi\ll 1$.

\item S. Weinberg, {\em Gravitation and Cosmology} (Wiley, New York, 1972), 
Eq.~(15.10.13).  In conformity with common present notation, the symbol $R(t)$ 
used in this book for the Robertson--Walker scale factor has been replaced in 
the present paper with $a(t)$.

\item S. Weinberg, ref. 7, Eq.~(15.10.52).

\item S. Weinberg, ref. 7, Eq.~(15.10.51).

\item S. Weinberg, ref. 7, Eq.~(15.10.53).  A factor $T$ was missing in the 
second term in the curly brackets in Eq.~(15.10.53), and has been suppplied 
here.

\item These formulas are obtained by comparing the acoustic damping rate 
calculated by N. Kaiser,  Mon. Not. Roy. Astr. Soc. {\bf 202}, 1169 (1983) with 
the damping rate calculated for arbitrary values of $\eta$ and $\chi$ by S. 
Weinberg,  Ap. J. {\bf 168}, 175 (1971), Eq.~(4.15).  The latter article also 
gives values for $\chi$ and $\eta$, repeated in ref. 7: it gives the same value 
of $\chi$ and a value for $\eta$ that is $3/4$ the value quoted in Eq.~(55), but 
these results were based on calculations of L. H. Thomas, Quart. J. Math. 
(Oxford) {\bf 1}, 239 (1930), that assumed  isotropic scattering and ignored 
photon polarization.  (The same value for $\eta$ had been given by C. Misner, 
Ap. J. {\bf 151}, 431 (1968).) Kaiser's results are calculated using the correct 
differential cross section for Thomson scattering and take photon polarization 
into account, and therefore supersede the earlier value quoted for $\eta$.
As late as 1995 the wrong value of the damping rate was still being used, for 
instance by Hu and Sugiyama, ref. 5, but the correct rate was used by Hu and 
White, Ap. J. {\bf 479}, 568 (1997).

 \item J. A. Peacock, {\em Cosmological Physics} (Cambridge University Press, 
Cambridge, UK 1999), p. 591.

\item The derivation is given
for instance in S. Weinberg, ref. 7,  Eq.~(15.9.13).  The presence of the second 
term on the left-hand side  has as a consequence the well known decay $\propto 
1/a(t)$ of the peculiar velocities of non-relativistic free particles.  The 
factor $1/a(t)$ multiplying the gradient of the potential  enters  again to 
convert a derivative with respect to co-moving coordinates into a derivative 
with respect to coordinates that measure proper distances.

\item Hu and Sugiyama, ref. 5; Hu and White, ref. 11.

\item A. Dimitropoulos and L. P. Grishchuk, gr-qc/001087.

\item See, e. g., P. H. Frampton, Y. J. Ng, and R. Rohm, Mod. Phys. Lett. 13, 
2541 (1998).

\item W. Hu and M. White,  ref. 11.

\item R. K. Sachs and A. M. Wolfe, Ap. J. {\bf 1}, 73 (1967).

 \item S. Weinberg, ref. 7, Eqs.~(15.10.29) and (15.1.19).  (A misprint has been 
corrected 
here: the equals sign in the first line of Eq.~(15.10.29) has been changed to a 
minus sign.)

\item L. A. Kofman and A. A. Starobinskii, Sov. Astron. Lett. {\bf 11}, 271 
(1985).

\item The $1/\ell$ dependence was found in reference [20], but without 
consideration of a possible interference between this effect and the Doppler 
shift and intrinsic temperature shift.
\end{enumerate}
\end{document}